\documentclass[amsmath,twocolumn,showpacs,amssymb,aps,nofootinbib]{revtex4-1}
\usepackage{graphicx}
\usepackage{subcaption}
\usepackage{pgfplots}
\usepackage{dcolumn}
\usepackage{bm}
\usepackage{slashed}
\usepackage{amsmath}
\usepackage{amssymb}
\usepackage{tikz}
\usepackage{bbold}
\usepackage{mathtools}
\usepackage{romannum}
\usepackage{natbib}
\usepackage{braket}
\begin{document}
\pagenumbering{arabic}
\title{Nielsen complexity of coherent spin state operators \\}
\author{Kunal Pal}
\email{kunalpal@iitk.ac.in}	
\author{Kuntal Pal}
\email{kuntal@iitk.ac.in}
\author{Tapobrata Sarkar}
\email{tapo@iitk.ac.in}
\affiliation{
Department of Physics, Indian Institute of Technology Kanpur, Kanpur 208016, India}
\date{\today}

\begin{abstract}
We calculate Nielsen's circuit complexity of coherent spin state operators. An expression for the complexity is obtained 
by using the small angle approximation of the  Euler angle parametrisation of a general $SO(3)$ rotation. This is then
extended to arbitrary times for systems whose time evolutions are generated by couplings to an external field, 
as well as non-linearly interacting Hamiltonians. In particular, we show how the 
Nielsen complexity relates to squeezing parameters of the one-axis twisted Hamiltonians in a transverse field, thus indicating its 
relation with pairwise entanglement. We further point out the difficulty with this approach for the Lipkin-Meshkov-Glick model, 
and resolve the problem by computing the complexity in the Tait-Bryan parametrisation.

\end{abstract}
\maketitle

\section{Introduction}
\label{intro}

In recent years, the notion of complexity has been popular in the study of quantum systems. 
Broadly speaking, complexity of a system measures the difficulty in preparing a target state, starting from
a given reference state. A standard approach of measuring complexity is the one pioneered by 
Nielsen \cite{Nielsen1, Nielsen2, Nielsen3}. In this geometric approach, the difficulty in constructing
a unitary operator that relates the target state to the reference is quantified by a cost functional 
for various paths in the space of unitary transformations. To be physically reasonable, these cost
functionals have to satisfy certain conditions, and then these can be shown to define length functionals 
on a Finslerian manifold, one that reduces to a Riemannian manifold for a specific choice of cost functional. 
Finding the complexity then reduces to a variational problem of finding geodesics on the space of unitary 
transformations. 

The above discussion can be quantified by the following mathematical steps, which have been outlined
in the more recent works of \cite{Myers1,BSS,Myers2,ABHKM}. First, given a reference state $|\Psi^R\rangle$, we
need a unitary operator $U(\tau)$, such that the target state is given as 
$|\Psi^T(\tau=1)\rangle= U(\tau)|\Psi^R(\tau=0)\rangle$, where $\tau$ parametrises a path 
in the Hilbert space, with its endpoints $\tau=0$ and $\tau=1$ denoting the reference and target
states, respectively. Next, we represent $U(\tau)$ as a path ordered exponential by expanding the
Hamiltonian generator in a given basis, 
\begin{equation}
U(\tau)=\overleftarrow{\cal P}\exp\left[\int_{0}^{\tau}d\tau'\sum_{a}Y^{a}(\tau')J_{a}\right]~,
\label{pathorder}
\end{equation}
where, for our purposes, $J_a$ will denote the generators of the $SO(3)$ group  and the coefficients $Y^{a}$ are 
called the control functions. Now, by differentiating Eq. (\ref{pathorder}) and using the orthogonality
relation between the $SO(3)$ generators, one obtains 
\begin{equation}
Y^a =\frac{1}{{\rm Tr} \Big[J^{a}.J_{a}^{T}\Big]} {\rm Tr}\Big(\partial_{\tau}U\cdot U^{-1} \cdot (J^a)^{T}\Big)~,
\end{equation}
where from now on, repeated indices will imply summation over these, in the usual sense. 
Then, we will define a length functional which makes the geometric
analysis clear \cite{Nielsen1}, \cite{Myers1}
\begin{equation}
{\mathcal C}(U) = \int_0^1d\tau\sqrt{G_{ab}Y^a(\tau)Y^b(\tau)}= \int_0^1 ds~.
\label{CU}
\end{equation}
The minimum value of the functional defined in Eq. (\ref{CU}) gives the required complexity. 
In order to proceed further, a metric on the
space of unitary transformations is obtained from 
\begin{eqnarray}
ds^{2} &=& G_{ab}\Big[\frac{1}{{\rm Tr} \big[J^{a}.J_{a}^{T}\big]}\text{Tr}
\Big(\text{d}U(\tau)\cdot U^{-1}(\tau)\cdot J_{a}^{T}\Big)
\Big]\times\nonumber\\ &~&\Big[\frac{1}{{\rm Tr} \big[J^{b}.J_{b}^{T}\big]}\text{Tr}
\Big(\text{d}U(\tau)\cdot U^{-1}(\tau)\cdot J_{b}^{T}\Big)\Big]^{*}~,
\label{met}
\end{eqnarray}
where for simplicity, we choose $G_{ab} = 4\delta_{ab}$. 
Now, this is equivalently written as $ds^{2}= g_{ij}dy^idy^j$, with $y^i$ denoting a
set of coordinates on the space of unitaries. 
Then, the length functional can be written as 
\begin{equation}\label{functional}
\mathcal{C}\big(U\big)=\int_{0}^{1}\text{d}\tau \sqrt{g_{ij}\frac{dy^{i}(\tau)}{d\tau}\frac{dy^{j}(\tau)}{d\tau}}~,
\end{equation}
and minimising this amounts to finding the geodesics on the space of unitary transformations. Here, the
boundary condition that encodes the information of the reference (target) state is at $\tau=0$ ($\tau=1$). 
On geodesics, the quantity $K=\sqrt{g_{ij}\frac{dy^{i}
(\tau)}{d\tau}\frac{dy^{j}(\tau)}{d\tau}}$
is a constant, and $K$, computed with appropriate boundary conditions at $\tau = 0$ and $\tau = 1$ is then
the measure of the Nielsen complexity ${\mathcal C}(U)$, as follows from Eq. (\ref{functional}). 

This approach has been immensely popular of late in the context of the holographic
principle, where the gauge-gravity duality relates strongly coupled quantum field theories
to weakly coupled gravity. Complexity in spin systems have also been well 
studied in the literature. In particular, \cite{Paraj, tapo1}  studied Nielsen complexity in
the context of quantum phase transitions in the Kitaev model and the transverse field
XY model, respectively. These works indicate that circuit complexity is a strong indicator of
such phase transitions, i.e., it becomes non-analytic at critical lines (see also \cite{xiong}). The work of
\cite{tapo1} also explored a different definition of complexity, namely the Fubini-Study 
complexity \cite{FS}, arising out of the definition of the quantum information metric \cite{ProvostVallee}, 
and established that this also shows special properties across a quantum
phase transition. 

What is relatively unexplored during the current flurry of activities in the study of complexity, 
is its behaviour in statistical and spin systems away from
phase transitions. Indeed, such limits often show interesting behaviour and are useful 
in understanding many of the founding principles of quantum mechanics. These
are important in their own right, and might throw light on the foundational issues of quantum mechanics. 
Fortunately, several methods developed in the context of the holographic principle give us a precise mathematical
formalism to tackle such problems. 

In this spirit, in the present work we study Nielsen complexity of operators corresponding to coherent spin states 
(CSS) \cite{JR,AEGT,JPG,AP}. We note here that circuit complexity associated with coherent states 
have been studied recently in the literature. In \cite{Myers2}, the authors calculated the complexity of 
bosonic coherent states, in \cite{RQY} the complexity between two coherent states was studied (see also \cite{GT}), 
and the complexity of generalised coherent states was obtained in \cite{GFJLC} using the covariance matrix 
method of calculating circuit complexity developed in \cite{CEHHJMM}.
Coherent states are 
one of the most fundamental objects in quantum physics, and  the closest cousins to classical
states since the standard harmonic oscillator coherent states as well as the CSS  saturate the corresponding uncertainty bounds. These have now been studied for many decades,
and have found many applications. For example, bosonic squeezed states, which are obtained by 
acting a non-linear operator on a bosonic coherent state is a widely studied subject 
in quantum optics. Similarly, squeezed spin states are also extremely well studied in the literature
and are known to give rise to pairwise entanglement \cite{Wang} (for further details, 
see the review \cite{Review}). 

Here, we study circuit complexity of the infinitesimal form of coherent spin state operators. 
We first review the basic details 
in the next section, and show that the Euler angle representation provides a convenient way
to compute the complexity. Next, in section \ref{sec3}, we present our main formula for the
complexity of these states, followed by a generalisation to arbitrary times  in 
section \ref{sec4}.  There, we apply the formalism to models well studied, both theoretically and 
experimentally in the context of spin squeezing, and 
explicitly show how complexity is related to the squeezing parameter of the initial state. Next, in section 
\ref{sec5}, we argue that the Euler angle representation is not always useful, and 
compute the complexity in the Tait-Bryan parametrisation and exemplify this  
in the context of the Lipkin-Meshkov-Glick model. 

\section{The coherent spin states}
\label{sec2}

The CSS (or Bloch state) can be constructed by applying the unitary operator 
(which we will call $\mathcal{D}^{j}$) on a normalised Hilbert space \cite{AP, JPG}. 
We want to calculate the circuit complexity of creating the operator $\mathcal{D}^{j}$,
starting from the unity operator. This is related  to the circuit complexity of the CSS (here the target state) starting from a state 
of the Hilbert space (the reference state) in the following way.  As we want to construct the state closest to the 
classical one, we take the reference state to be lowest or the highest weight state. 
Then we have to calculate the complexity associated with all the unitary operators that connect 
these two states and the minimum value among these is the circuit complexity of the CSS \cite{YK,YANZK,BS}.

We start by considering the angular momentum operator in three dimensional space, collectively denoted by the 
vector $\mathbf{J}=\{\mathbf{J}_{x},\mathbf{J}_{y},\mathbf{J}_{z}\}$. They satisfy the usual commutation relations 
$\big[\mathbf{J}_{i},\mathbf{J}_{j}\big]=i\mathbf{J}_{k}$ with $i,j,k=x,y,z$.
The operators $\mathbf{J}_{\pm}=\mathbf{J}_{x}\pm i\mathbf{J}_{y}$ and $\mathbf{J}_{3}$ satisfy the Lie algebra  of the group $SO(3)$.
Thus if we denote a basis element of the Hilbert space  
as $\big|j,m\big>$, with $j$ being the eigenvalue of the Casimir operator 
$\mathbf{J}$ and $m$ the eigenvalue of $\mathbf{J}_{3}$, then the result  of acting the operators 
$\mathbf{J}_{\pm}$ and $\mathbf{J}_{3}$ on them is the same that of the action of a unitary irreducible representation 
of the three dimensional rotation group for an infinitesimal rotation. Now since a general rotation in three dimensional 
space can be represented by the axis angle representation, as well as in terms of the Euler angles \cite{WT}, we can 
write the operator  $\mathcal{D}^{j}$ in either of the two representations. However as we  shall see, 
Euler angles are easier to use in the calculation of circuit complexity, because in terms of Euler angles a general 
rotation can be written in terms of three rotations about two fixed axes, so that the matrix form of a general unitary 
operator $U$ associated with the rotation group is simpler. Below we shall write the matrix form of operator  
$\mathcal{D}^{j}$ and the associated CSS in both representations.

\subsection{The axis-angle representation}

In the axis-angle representation, a general rotation in three dimensional space is specified by 
the unit vector $\mathbf{n}$ along the axis about which the rotation is performed, and an angle $\theta$
which gives the amount of rotation around $\mathbf{n}$. 
The unitary operator associated with this rotation is
$\mathcal{D}(\mathcal{R}(\mathbf{n},\theta))=\exp\big[-i\theta \mathbf{n}\cdot\mathbf{J}\big]$. 
To construct a CSS in this representation, let us consider a unit vector $\mathbf{n}_{\phi}$ in the $xy$ plane, making 
an angle $\phi$ ($0\leq\phi\leq2\pi$) with the $y$ axis, so that its components are 
$\mathbf{n}_{\phi}=\{-\sin \phi, \cos \phi,0 \}$. We are interested in the particular rotation about this 
axis, which brings the unit vector along the $z$ direction ($\textbf{z}=(0,0,1)$) to an arbitrary unit 
vector $\textbf{r}(\theta,\phi)$ for our three dimensional space. 
The corresponding unitary operator is given by
\begin{eqnarray}
\label{diaxa}
\mathcal{D}^{j}(\mathcal{R}_{\mathbf{r}}) &=&
\exp\big[-i\theta \big(-\sin \phi \mathbf{J}_{x}+\cos \phi \mathbf{J}_{y}\big)\big]\nonumber\\
&=&\exp\big[\xi \mathbf{J}_{+}-\bar{\xi}\mathbf{J}_{-}\big]~~,
\end{eqnarray}
where the complex number $\xi$ is defined by $\xi=-\frac{\theta}{2}\exp\big[-i\phi\big]$, and an overhead bar 
denotes its complex conjugate. Now consider an arbitrary normalised state $\big|\Psi_{0}\big>=
\sum_{m=-j}^{j}c_{m}\big|j,m\big>$ with $\sum_{m=-j}^{j}\big|c_{m}\big|^{2}=1$.  The CSS associated with $\big|\Psi_{0}\big>$ is obtained by acting it with the unitary operator $\mathcal{D}^{j}(\mathcal{R}_{\mathbf{r}})$ i.e. $\big|\mathbf{r}\big>=\big|\theta,\phi\big>=\mathcal{D}^{j}(\mathcal{R}_{\mathbf{r}})\big|\Psi_{0}\big>$. 

To construct a generalised coherent state, we can act the operator $\mathcal{D}^{j}(\mathcal{R}_{\mathbf{r}})$  
on any arbitrary (normalised) linear combination of the base states. However, as is well known \cite{AP}, the CSS 
which has properties closest to the classical case is obtained when one chooses  
$\big|\Psi_{0}\big>$ to be the  highest weight state  $\big|j,j\big>$ or the lowest weight state  $\big|j,-j\big>$. The standard expressions for the CSS in this case can  be written in terms of the angles $\theta,\phi$ as a linear combination of states   $\big|j,m\big>$, and is given by \cite{JPG,JR,AEGT}
\begin{eqnarray}
\label{CSS}
\big|\mathbf{r}\big>&=&
\sum_{m=-j}^{j}\sqrt{\left(
\begin{array}{ccc}
2j  \\
j+m\\
\end{array}
\right)}\Big(\cos \frac{\theta}{2}\Big)^{j+m}\Big(\sin \frac{\theta}{2}\Big)^{j-m}\times\nonumber\\
&~&\exp
\Big[i\big(j-m\big)\phi\Big]\big|j,m\big>~.
\end{eqnarray}

As mentioned above, the circuit complexity of preparing the target state
$\big|\mathbf{r}\big>$ starting from the reference state $\big|j,j\big>$ (or $\big|j,-j\big>$)
is related  to the complexity of obtaining  the  unitary operator $\mathcal{D}^{j}(\mathcal{R}_{\mathbf{r}})$ starting from 
the identity operator $\mathbf{I}$. To calculate the latter in Nielsen's geometric approach,  we need a  matrix representation of the 
linearised version of  $\mathcal{D}^{j}(\mathcal{R}_{\mathbf{r}})$, written in terms of a suitable basis.

To this end, note that we can write \cite{JPG,AEGT}
$\mathcal{D}^{j}(\mathcal{R}_{\mathbf{r}})
=\exp\big[-\bar{\eta}
\mathbf{J}_{+}\big]\exp\big[\ln \big(1+|\eta|^{2}\big)\mathbf{J}_{z}\big]\exp\big[\eta \mathbf{J}_{-}\big]$,
where $\eta=\tan\frac{\theta}{2}\exp\big[i\phi\big]$. To write down the matrix form of $\mathcal{D}^{j}
(\mathcal{R}_{\mathbf{r}})$, we use the standard $3\times3$ matrix representation of the generators 
$\mathbf{J}_{i}$  of  rotation \cite{WT} to get  
\begin{equation}\label{aadis}
\mathcal{D}^{j}\big(\mathcal{R}_{\mathbf{r}}(\eta)\big)=\left(
\begin{array}{ccc}
1 &-if(\eta) & \bar{\eta}+\eta  \\
if(\eta) & 1 &i\big(\bar{\eta}-\eta\big) \\ 
-\big(\bar{\eta}+\eta\big) & i\big(\eta-\bar{\eta}\big) & 1\\
\end{array}
\right)~,
\end{equation}
where $f(\eta)=\ln \big(1+|\eta|^{2}\big)$.

\subsection{The Euler angles} 

In terms of the three Euler angles ($\alpha,\beta,\gamma$), the general rotation operator 
$\mathcal{R}(\alpha,\beta,\gamma)$ is given by the product of three rotations performed in the order \cite{WT,SN}
$\mathcal{R}(\alpha,\beta,\gamma)=\mathcal{R}_{z}(\alpha)\mathcal{R}_{y}(\beta)\mathcal{R}_{z}(\gamma)$.
In the product, all the rotations are performed with respect to the fixed $z$, $y$ and $z$ axes respectively. 
As before, in the construction of the CSS, we are interested in a particular rotation, namely the one which brings 
the unit vector $\mathbf{z}$ along the $z$ axis to the position of the unit vector $\textbf{r}(\theta,\phi)$ 
in spherical polar coordinates. The required rotation is given by the operator  
$\mathcal{R}(\phi,\theta,0)=\mathcal{R}_{z}(\phi)\mathcal{R}_{y}(\theta)$. Using the same matrix form of the 
generators as the ones used in Eq. (\ref{aadis}), we now have the following simplified  matrix form of 
this operator, up to first order in rotation angles,
\begin{equation}\label{Eadis}
\mathcal{D}^{j}\big(\mathcal{R}_{\mathbf{r}}(\theta,\phi)\big)=\left(
\begin{array}{ccc}
1 &-\phi & \theta  \\
\phi & 1 &0 \\ 
-\theta & 0 & 1\\
\end{array}
\right)~.
\end{equation}
Comparing with Eq.  (\ref{aadis}), this form is easier to use. So is the form of the general unitary operator 
which acts on the states of the Hilbert space and is given in a standard fashion by 
$U(\alpha,\beta,\gamma)=\exp\Big(-i\alpha \mathbf{J}_{z}\Big)\exp\Big(-i\beta \mathbf{J}_{y}\Big)\exp\Big(-i\gamma \mathbf{J}_{z}\Big)$.
For infinitesimal rotations, the matrix form of this operator can be calculated to be 
\begin{equation}\label{EAUn}
U(\alpha,\beta,\gamma)=\left(\begin{array}{ccc}
1 &-\big(\alpha+\gamma\big) & \beta \\
\big(\alpha+\gamma\big) & 1 &0 \\ 
-\beta & 0 & 1\\
\end{array}
\right)\equiv \left(\begin{array}{ccc}
1 &-\gamma & \beta \\
\gamma & 1 &0 \\ 
-\beta & 0 & 1\\
\end{array}
\right)~,
\end{equation}
where, in the last line, we have redefined the angle $\gamma$. As can be seen from the last equation, in the 
infinitesimal version of the general rotation, only the combination of two rotations (by angles $\alpha$ and $\gamma$) 
around the fixed $z$ axis appears, so that the redefinition in the last expression does not affect the calculations below.

Before ending this section, we note that here we have used the infinitesimal form of the rotation operator, and 
this will be used to find the metric on the space of unitaries. Since the metric is a local quantity, it indeed
suffices to work with this infinitesimal form. We will keep this in mind in our analysis below. 

\section{Circuit complexity in the Euler angle representation}
\label{sec3}

To calculate the circuit complexity of forming the operator  
$\mathcal{D}^{j}\big(\mathcal{R}_{\mathbf{r}}(\theta,\phi)\big)$ given in Eq.
(\ref{Eadis}) starting from the identity 
operator $\mathbf{I}$, we need to find out the geodesic connecting these two unitary operators in the space of 
unitary operators $U$. The minimum length of such a geodesic is the required circuit complexity. As mentioned
in the introduction, This 
procedure of calculating the circuit complexity using  Nielsen's geometric approach is by now 
standard, and more details can be found in \cite{Myers1,Myers2,BSS,BCHMY}. We shall only 
outline  the procedure briefly. To describe a geodesic trajectory on the space of unitary operators, 
we use the parameter $\tau$ along the trajectory, with the parameter is chosen in such a way that at the 
starting point of the geodesic we have $\tau=0$ and at the ending $\tau=1$. Hence the boundary 
conditions at the two points are respectively
\begin{equation}
U(\tau=0)=\mathbf{I}~;~~U(\tau=1)=\mathcal{D}^{j}\big(\mathcal{R}_{\mathbf{r}}(\theta,\phi)\big)~.
\end{equation}
Comparing these boundary conditions with the infinitesimal form of unitary operator of the rotation group in Eq. (\ref{EAUn}),
we can express them in terms of $\beta,\gamma$ at the two points to be
\begin{eqnarray}
\label{bcgb}
\beta,\gamma\big(\tau=0\big)=0~,
\beta\big(\tau=1\big)=\theta~,~\gamma\big(\tau=1\big)=\phi~.
\end{eqnarray}

After setting the boundary conditions, the next step is to find out the metric on the space of unitaries, 
given by Eq. (\ref{met}). 
Writing $y_{i}=\{\beta,\gamma\}$, and taking $G_{ab}=4\delta_{ab}$ for convenience, the line element  $ds^{2}=g_{ij}dy^{i}dy^{j}$ is given by
\begin{equation}\label{bgmetric}
\begin{split}
ds^{2}=\frac{1}{\big(1+\beta^{2}+\gamma^{2}\big)^{2}}\Bigg[\Big(\gamma^{4}+\gamma^{2}
\big(\beta^{2}+5\big)+4\Big)\text{d}\beta^{2} \\
+\Big(\beta^{4}+\beta^{2}\big(\gamma^{2}+5\big)+4\Big)\text{d}\gamma^{2}-
2\beta\gamma\Big(\beta^{2}+\gamma^{2}+5\Big)\text{d}\beta\text{d}\gamma\Bigg]~.
\end{split}
\end{equation}

This form of the metric is non-diagonal, and therefore cumbersome to deal with. However, there is 
a hidden symmetry here. To see this, we introduce two new 
coordinates $\{\rho,\Theta\}$  related to $\{\beta,\gamma\}$ by 
\begin{equation}
\beta\big(\rho,\Theta\big)=\rho\sin\Theta~,~~\gamma\big(\rho,\Theta\big)=\rho\cos\Theta~.
\label{bcin}
\end{equation}
Substituting $\text{d}\beta\big(\rho,\Theta\big)$ and $\text{d}\gamma\big(\rho,\Theta\big)$ in Eq. (\ref{bgmetric}),
we have the following simplified diagonal form of the metric,
\begin{equation}\label{diagonal}
ds^{2}=\frac{4}{\big(1+\rho^{2}\big)^{2}}\text{d}\rho^{2}+\frac{\rho^{2}\big(4+\rho^2\big)}{\big(1+\rho^2\big)}
\text{d}\Theta^{2}~.
\end{equation}
In terms of our new coordinates, the metric has not only become diagonal, but $\Theta$ has also become cyclic, indicating
the hidden symmetry. This is indeed a big advantage, as we will see. Before proceeding, we need to write the boundary conditions 
in Eq. (\ref{bcgb}) in the $\{\rho,\Theta\}$ coordinates. These are given by
\begin{eqnarray}
\label{bcrt}
&~&\rho\big(\tau=0\big)=0~,~
\rho\big(\tau=1\big)=\sqrt{\theta^{2}+\phi^{2}}~,~\nonumber\\
&~&\Theta\big(\tau=1\big)=\arctan\big[\theta/\phi\big]~,
\end{eqnarray}
with the value of $\Theta(\tau=0)$ being indeterminate. 

To calculate the circuit complexity, we minimise the complexity functional in Eq. (\ref{functional}) 
with the coordinates $y^{i}=\{\rho,\Theta\}$. This is done by solving the geodesic equations 
subjected to boundary conditions in Eq. (\ref{bcrt}), in the space of unitaries 
equipped with the metric of Eq. (\ref{diagonal}). The geodesic equations
are equivalent to the ones obtained from the Lagrangian 
$\mathcal{L}=g_{ij}\dot{y}^{i}\dot{y}^{j}$, where an overdot indicates a derivative with respect to $\tau$.
We find that these are given by
\begin{eqnarray}
&~&{\ddot \rho}-\frac{2 \rho {\dot\rho}^2}{\rho^2+1}-\frac{1}{4} \rho \left(\rho^4+2 \rho^2+4\right) {\dot\Theta}^2=0~,\nonumber\\
&~&{\ddot\Theta}+\frac{2 \left(\rho^4+2 \rho^2+4\right) {\dot\rho}}{\rho \left(\rho^2+1\right) \left(\rho^2+4\right)} {\dot\Theta}=0~.
\label{geoeqs}
\end{eqnarray}
From Eq. (\ref{geoeqs}), we see that $\Theta = {\rm constant}$ are geodesics. Then, from the first equation 
in Eq. (\ref{geoeqs}), we obtain with $g_{ij}\dot{y}^{i}\dot{y}^{j}=K^2$,
\begin{equation}
\dot{\rho}=\frac{K}{2}\big(1+\rho^{2}\big)~,
\end{equation}
a conclusion that is also reached from the definition of $K^2$, along with the fact that $\Theta$ is a constant. 
We then have the solutions,
\begin{equation}
\Theta=\Theta_{0}~,~~\rho\big(\tau\big)=\tan\Big[\frac{K\tau}{2}-C_{1}\Big]~,
\label{sols}
\end{equation}
where $\Theta_0$ and $C_1$ are two constants that have to be fixed from boundary conditions. Note that
we must have $\Theta_0\neq 0$, as otherwise Eq. (\ref{bcin}) would imply that $\beta = 0$ for all values
of $\tau$, a condition that is clearly incompatible with that at $\tau = 1$ in Eq. (\ref{bcgb}).  

Now, translating back to the $\{\beta,\gamma\}$ coordinates, we have 
\begin{equation}
\beta = \tan\Big[\frac{K\tau}{2}-C_{1}\Big]\sin\Theta_0~,~
\gamma = \tan\Big[\frac{K\tau}{2}-C_{1}\Big]\cos\Theta_0~.
\label{finalbc}
\end{equation}
Eq. (\ref{bcgb}) then determines that the constants $C_1=n\pi$ and $\Theta_0 = \arctan(\theta/\phi)$. Setting $\tau=1$, 
either of the two relations in Eq. (\ref{finalbc}) then gives the complexity 
\begin{equation}
\label{CSScom}
\mathcal{C}\Big(\mathcal{D}^{j}\big(\mathcal{R}_{\mathbf{r}}(\theta,\phi)\big)\Big)= K=
2\Big(\arctan\Big[\sqrt{\theta^{2}+\phi^{2}}\Big]+n\pi\Big)~.
\end{equation}
The metric in Eq. (\ref{bgmetric}) and the above expression for the circuit complexity  
is valid for small values of rotation angles $\theta,\phi$ such that we can neglect the $\mathcal{O}(\theta\phi)$ 
and higher order terms in the  operator $\mathcal{D}^{j}\big(\mathcal{R}_{\mathbf{r}}(\theta,\phi)\big)$ and  
the matrix form of the operators given in Eqs. (\ref{Eadis}) and (\ref{EAUn}) provide good approximation to the exact expressions.

As a simple application of Eq. (\ref{CSScom}), we can compute the complexity of the Dicke model coherent states,
which can be written as the tensor product of the coherent state of the harmonic oscillator 
and the coherent spin state \cite{AFLN}
\begin{equation}
\big|\alpha,\mathbf{r}\big>=\big|\alpha\big> \otimes \big|\mathbf{r}\big>~~.
\end{equation}
Here $\alpha$ is a complex number which is the eigen value of the annihilation operator $
a\big|\alpha\big>=\alpha\big|\alpha\big>$. The state $\big|\alpha\big>$ can be obtained from the 
unit mass harmonic oscillator ground state by applying the displacement operator $\mathcal{D}(\alpha)$ i.e.
\begin{equation}
\big|\alpha\big>=\mathcal{D}(\alpha)\big|0\big>~,~~\mathcal{D}(\alpha)=
\exp\Big[i\sqrt{2}\Big(\sqrt{\omega}\alpha_{i}\mathbf{Q}-\frac{\alpha_{r}\mathbf{P}}{\sqrt{\omega}}\Big)\Big]~~~.
\end{equation}
where $\mathbf{Q}$ and $\mathbf{P}$ are the operator form of the position and momenta respectively,
and $\alpha_{r},\alpha_{i}$ are the real and imaginary parts of $\alpha$. 
The complexity of creating the Dicke model CS ($\big|\alpha,r\big>$) from the product state 
($\big|0\big>\otimes \big|j,-j\big>$) can be viewed as the sum of complexities of creating the 
displacement operator  $\mathcal{D}(\alpha)$ and the rotation operator $\mathcal{D}^{j}(\mathcal{R}_{\mathbf{r}})$ 
from the identity (see \cite{YK,YANZK}).
It follows straightforwardly that 
\begin{eqnarray}
\mathcal{C}_{Dicke}&=&\mathcal{C}\Big(\mathcal{D}\big(\alpha)\Big)+\mathcal{C}
\Big(\mathcal{D}^{j}\big(\mathcal{R}_{\mathbf{r}}\big)\Big)
=\sqrt{2\Big(\frac{\alpha_{r}^{2}}{\omega } +\omega \alpha_{i}^{2}\Big)}\nonumber\\
&+& 2\Big(\arctan
\Big[\sqrt{\theta^{2}+\phi^{2}}\Big]+n\pi\Big)~.
\end{eqnarray}
Here the the first term $\mathcal{C}\big(\mathcal{D}\big(\alpha)\big)$ has been  calculated recently in \cite{BCHMY}.\footnote{In \cite{BCHMY} the complexity of displacement operator of the inverted harmonic oscillator was calculated. However the result can be easily generalized for harmonic oscillator as well.} 

\section{Time evolution of Circuit complexity }
\label{sec4}

So far we have computed the circuit complexity associated with CSS at a fixed time. 
It is more interesting to compute this in the case where the system evolves in time, with the  
time evolution being governed by the Hamiltonian operator $\mathbf{H}$. 
The procedure of calculating the complexity remains the same as in previous section, 
the only thing that is different is the expressions of the matrix 
$\mathcal{D}^{j}\big(\mathcal{R}_{\mathbf{r}}(\theta,\phi)\big)$ (see Eq. (\ref{Eadis})), 
which now becomes  a function of time as well, and subsequently the boundary 
conditions of Eq. (\ref{bcgb}) are also changed. 

At an arbitrary time $t$, we have
\begin{equation}\label{Eadis-t}
\mathcal{D}^{j}\big(\theta,\phi,t\big)=\left(
\begin{array}{ccc}
1 &-f(\theta,\phi,t) & g(\theta,\phi,t) \\
f(\theta,\phi,t) & 1 &0 \\ 
-g(\theta,\phi,t) & 0 & 1\\
\end{array}
\right)~,
\end{equation}
where $f(\theta,\phi,t),g(\theta,\phi,t)$ are two functions whose forms depend on the Hamiltonian of the system. 
The modified boundary conditions at the starting and the end point of the geodesic now read 
\begin{equation}\label{bcgbt}
\begin{split}
&\beta,\gamma\big(\tau=0\big)=0~,~~
\beta\big(\tau=1\big)=g(\theta,\phi,t)~,~~\\
&\gamma\big(\tau=1\big)=f(\theta,\phi,t)~.
\end{split}
\end{equation}
The calculation of the circuit complexity is the same as before, and we 
will write down the final expression only,  
\begin{equation}\label{com-t}
\mathcal{C}\big(\theta,\phi,t\big)= 2\arctan\Big[\sqrt{f(\theta,\phi,t)^{2}+g(\theta,\phi,t)^{2}}\Big]+2n\pi~.
\end{equation}
We shall now consider three different systems having Hamiltonian with increasing intricacy, 
and we will see the time evolved operator $\mathcal{D}^{j}\big(\theta,\phi,t\big)$ for each of these 
systems will fall into a distinct category depending upon  the spin components present. 
Namely, in the standard Euler angle representation of Eq. (\ref{EAUn}), rotations
about only two fixed axes appear (here $x$ and $z$ axes). In the first two
examples below, at time $t$, only rotations about these two axes appear with the
difference between them is that in the first case, rotation about the third
axis (here $x$) is absent in any order of rotation angle while in the
second case the third axis rotation is absent only in the lowest order of
the rotation angles.  In the third case however, in
$\mathcal{D}^{j}\big(\theta,\phi,t\big)$ a rotation about the $x$
axis is present even in the lowest order in $\theta,\phi$, and hence to
obtain an analytic expression for the complexity, we have to switch to the alternative Tait-Bryan
parametrisation for the rotation matrix.

\subsection{Class-1 : Spin magnet interaction}
\label{class1}

First we consider the simple case of a spin $\mathbf{S}$ interacting with a constant external magnetic field
$\mathbf{B}$ along the $z$ direction  via an interaction Hamiltonian given by $-\mathbf{S} \cdot 
\mathbf{B}=-~\mathbf{S}_{z}B$.
In this case, it is straightforwardly shown that 
at any time $t$ of the evolution, the CSS operator 
remains coherent (which is not true for two other examples we consider below) and hence the rotation operator at 
$t$ can be obtained by replacing $\phi$ by $\phi+Bt$,
in the final rotation about the $z$ axis. Now taking the linear form of this operator for small values of rotation 
angles, we find that the complexity at time $t$ is
\begin{equation}
\label{class1eq}
\mathcal{C}\big(\theta,\phi,t\big)=2\arctan\bigg[\sqrt{\theta^{2}
+\big(\phi+Bt\big)^{2}}\bigg]+2n\pi~.
\end{equation}
This result follows from standard textbook material with the only non trivial  
point being the transformation between the Euler angle and the axis angle representations. 
The necessary  details are summarised in Appendix \ref{AppendixA}.
For given non-zero values of $\theta$ and $\phi$, the complexity thus monotonically 
increases and saturates to a maximum value of $\pi$ at large times. 

\subsection{Class-2 : Non-linear interactions}
\label{class2}

Another interesting application of the results derived so far can be envisaged via non-lineaer interactions.
We consider a collection of $N$ spins and assume that there is a nonlinear interaction present between the individual 
spins. We also assume that this nonlinear interaction term is of the form $\mathbf{J}_{i}^{2}$ where $i=x,y,z$. 
The resulting model is known in the literature as the one axis twisting model \cite{KU,LNL}.\footnote{If the nonlinear 
interaction term between  the spins is of the form $\mathbf{J}_{i}\mathbf{J}_{j}+\mathbf{J}_{j}\mathbf{J}_{i}$ 
with $i\neq j$ and $\mathbf{J}_{i}$ denoting $i$th component of collective angular momentum operator, then the model is known as  two axis counter twisting model \cite{KU}. Here we shall consider only the 
one axis twisting model. } Specifically, we are interested in is the one axis twisting model in
the presence of a transverse field, and the Hamiltonian reads
\begin{eqnarray}\label{Hssf}
\mathbf{H}_{ssf}=2\delta \mathbf{J}_{z}^{2}+\Omega \mathbf{J}_{x}~,
\end{eqnarray}
with the first term being the nonlinear interaction between the spins, and the second
represents an interaction with a transverse external field of frequency $\Omega$ \cite{LNL} (see also \cite{BSP}). 
As before, our aim is to find out the operator $\mathcal{D}^{j}\big(\theta,\phi,t\big)$ at a time $t$. 
For this, we need the  expressions of the operators $\mathbf{J}_{z}(t)$ and $\mathbf{J}_{y}(t)$. 

We start with the Heisenberg equation of motion for the spin operators, given respectively by \cite{LNL}
\begin{equation}
\begin{split}
& \frac{d\mathbf{J}_{x}(t)}{dt}=-4\delta \mathbf{J}_{(y}(t)\mathbf{J}_{z)}(t)~,~~ \\
& \frac{d\mathbf{J}_{y}(t)}{dt}=-\Omega \mathbf{J}_{z}(t) +4\delta \mathbf{J}_{(x}(t)\mathbf{J}_{z)}(t)~,~~\\
& \frac{d\mathbf{J}_{z}(t)}{dt}=\Omega \mathbf{J}_{y}(t)~.
\end{split}
\label{ssf}
\end{equation}
Here $\mathbf{J}_{(i}\mathbf{J}_{j)}=\frac{1}{2}\Big[\mathbf{J}_{i}\mathbf{J}_{j}+\mathbf{J}_{j}\mathbf{J}_{i}\Big]$ 
indicates symmetrization of the indices. The general solutions of these coupled first order equations are 
difficult to obtain, and they are usually solved by employing an approximation scheme. Here we shall use 
the so called frozen spin approximation, which is justified when the condition $\Omega>>\delta$ is satisfied. 
In that case, the force of the external field is much greater than the nonlinear interaction strength,
and hence $ \mathbf{J}_{x}(t)$ remains fixed at its value at $t=0$ during the entire evolution. 
This value of conveniently fixed to be $-J$ \cite{LNL}. With the frozen spin approximation, 
the solution to Eq. (\ref{ssf}) is given as (with the notation $\mathbf{J}_{i}=\mathbf{J}_{i}(t=0)$)
\begin{eqnarray}\label{sfs}
\begin{split}
&\mathbf{J}_{z}(t)\approx \mathbf{J}_{z}\cos\big[\omega_{0}t\big]+\frac{\Omega \mathbf{J}_{y}}
{\omega_{0}}\sin\big[\omega_{0}t\big]~~,~~\\
&\mathbf{J}_{y}(t)\approx -\frac{\omega_{0} \mathbf{J}_{z}}
{\Omega}\sin\big[\omega_{0}t\big]+\mathbf{J}_{y}\cos\big[\omega_{0}t\big]~,
\end{split}
\end{eqnarray}
where $\omega_{0}=\sqrt{\Omega^{2}+4\delta \Omega J}$ is known as frozen spin frequency. 
Hence at time $t$, we have 
\begin{equation}
\begin{split}
&\mathcal{D}^{j}\big(\theta,\phi,t\big)=e^{i\mathbf{H}_{ssf}t}\mathcal{D}^{j}\big(\theta,\phi,t=0\big)
e^{-i\mathbf{H}_{ssf}t}\\
&=\exp\Big(-i\phi \mathbf{J}_{z}(t)\Big)\exp\Big(-i\theta \mathbf{J}_{y}(t)\Big)~.
\end{split}
\end{equation}
Now we substitute the solutions of Eq. (\ref{sfs}) to get
\begin{equation}
\label{Dtssf}
\begin{split}
\mathcal{D}^{j}\big(\theta,\phi,t\big)&=1-i\mathbf{J}_{z}\left(\phi\cos\big[\omega_{0}t\big]-\frac{\theta \omega_{0}}
{\Omega}\sin\big[\omega_{0}t\big]\right)\\
& -i\mathbf{J}_{y}\left(\frac{\phi  \Omega}
{\omega_{0}}\sin\big[\omega_{0}t\big]+\theta\cos\big[\omega_{0}t\big]\right)+\mathcal{O}\big(\theta\phi\big)~.
\end{split}
\end{equation} 
Comparing with the previous example of the class-1 model, here no term proportional 
to $ \mathbf{J}_{x}$ appears in Eq. (\ref{Dtssf}). 
The reason for this is that in the previous case, the external field was applied along $\textbf{z}$ direction, 
but in contrast here it is along the $\textbf{x}$ direction. Since we are working in the frozen spin 
approximation, the time variation of $ \mathbf{J}_{x}$ is fixed by the external field.
Had we applied the field in $\textbf{z}$ or $\textbf{y}$ direction, such component would arise, as we 
shall soon see in the context of Lipkin-Meshkov-Glick model.

This expression for the operator $\mathcal{D}^{j}\big(\theta,\phi,t\big)$ in Eq. (\ref{Dtssf})
is a periodic function of time, and hence the 
complexity of creating such an operator from the identity also varies periodically with time.
This is in contrast to the spin-magnetic interaction that we studied before. 
As mentioned above, the reason for such behaviour has also to do with the direction of the applied field. 
As before, substituting the matrix form of the operator, and comparing with the general matrix operator of 
Eq. (\ref{Eadis-t}), we get the unknown functions in this case to be
\begin{equation}
\begin{split}
&f(\theta,\phi,t)=\Big(\phi\cos\big[\omega_{0}t\big]-\frac{\theta \omega_{0}}
{\Omega}\sin\big[\omega_{0}t\big]\Big)~,~~\\
&g(\theta,\phi,t)=\frac{\phi  \Omega}
{\omega_{0}}\sin\big[\omega_{0}t\big]+\theta\cos\big[\omega_{0}t\big]~.
\end{split}
\end{equation}
Thus the circuit complexity of the CSS is given by the expression
\onecolumngrid
\begin{equation}\label{C1at}
\mathcal{C}\big(\theta,\phi,t\big)=2\arctan\bigg[\sqrt{\left(\phi\cos\big[\omega_{0}t\big]-\frac{\theta \omega_{0}}
{\Omega}\sin\big[\omega_{0}t\big]\right)^{2}+\left(\frac{\phi  \Omega}
{\omega_{0}}\sin\big[\omega_{0}t\big]+\theta\cos\big[\omega_{0}t\big]\right)^{2}}\Bigg]+2n\pi~
\end{equation}
\twocolumngrid
Since we are working with the frozen spin approximation, if we take the limit $\delta \rightarrow 0$, i.e.,
$\omega_0 \to \Omega$ in this equation, we get back the time independent expression for the 
complexity of Eq. (\ref{CSScom}), as expected. 

Having derived the expression  for the circuit complexity, it is interesting to note that this can be written in 
terms of the squeezing parameters of the single axis twisting model. We first consider the unit vector 
$\textbf{n}$ having component $\textbf{n}_{i}$ with respect to a set of mutually orthogonal unit vectors. 
Then, for a many particle system which has the collective spin components  
$\textbf{J}_{\mathbf{n}}=\mathbf{n}\cdot\mathbf{J}$, we will, following \cite{SDCZ},\cite{Review},
define the squeezing parameter along a direction $\textbf{n}_{1}$ as
\begin{equation}
\xi^{2}_{\textbf{n}_{1}}=\frac{N\big<\big(\Delta \textbf{J}_{\textbf{n}_{1}}\big)^{2}\big> }
{\big<\textbf{J}_{\textbf{n}_{2}}\big>^{2}+\big<\textbf{J}_{\textbf{n}_{3}}\big>^{2}}~,
\end{equation}
and a state is said to be spin squeezed along the direction $\textbf{n}_{i}$ if $\xi^{2}_{\textbf{n}_{i}}<1$ for that state. 
It is well known that the applied external field increases squeezing compared to the zero external field case,
and that this squeezing can be maintained for a longer period of time as well \cite{LNL}.  Now we assume that 
at $t=0$ the single axis twisting model is prepared in an eigenstate of the operator $\textbf{J}_{x}$. 
Then  we can calculate the variance of the spin components in the frozen spin approximation 
\begin{equation}
\begin{split}
& \big<\big(\Delta \textbf{J}_{y}(t)\big)^{2}\big>=\frac{J}
{2}\Big(\cos^{2}\big[\omega_{0}t\big]+\frac{\omega_{0}^{2}}{\Omega^{2}}\sin^{2}\big[\omega_{0}t\big]\Big)~,\\
&\big<\big(\Delta \textbf{J}_{z}(t)\big)^{2}\big>=\frac{J}{2}\Big(\cos^{2}\big[\omega_{0}t\big]+\frac{\Omega^{2}}
{\omega_{0}^{2}}\sin^{2}\big[\omega_{0}t\big]\Big)~.
\end{split}
\label{sq1}
\end{equation}
We also calculate the correlation between the operators $\textbf{J}_{z}(t)$ and $\textbf{J}_{y}(t)$, 
which is given by 
\begin{equation}
\big<\textbf{J}_{z}(t)\textbf{J}_{y}(t)+\textbf{J}_{y}(t)\textbf{J}_{z}(t)\big>=J\cos\big[\omega_{0}t\big]
\sin\big[\omega_{0}t\big]\bigg(\frac{\Omega}{\omega_{0}}-\frac{\omega_{0}}{\Omega}\bigg)~.
\label{sq2}
\end{equation}
When the nonlinear interaction between the spins is zero, i.e., $\delta=0$ we have $\omega_{0}=\Omega$ and hence 
this correlation function vanishes as expected. Using Eqs. (\ref{sq1}) and (\ref{sq2}), 
we can write the final expression for the complexity obtained in Eq. (\ref{C1at}) as
\onecolumngrid 
\begin{equation}
\mathcal{C}\big(\theta,\phi,t\big)=\\
2\arctan\bigg[\sqrt{\frac{2}{J}\Big(\theta^{2}\big<
\big(\Delta \textbf{J}_{y}(t)\big)^{2}\big>+\phi^{2}\big<\big(\Delta \textbf{J}_{z}(t)\big)^{2}
\big>+\theta\phi\big<\textbf{J}_{z}(t)\textbf{J}_{y}(t)+\textbf{J}_{y}(t)\textbf{J}_{z}(t)
\big> \Big)}\Bigg]+2n\pi~.
\end{equation}
\twocolumngrid
In terms of the squeezing parameters introduced earlier, we have 
\onecolumngrid
\begin{equation}
\mathcal{C}\big(\theta,\phi,t\big)=2\arctan\bigg[\sqrt{\theta^{2}
\xi^{2}_{\textbf{n}_{y}}+\phi^{2}\xi^{2}_{\textbf{n}_{z}}
+\frac{2\theta\phi}{J}\big<\textbf{J}_{z}(t)\textbf{J}_{y}(t)+\textbf{J}_{y}(t)\textbf{J}_{z}(t)\big> }\Bigg]+2n\pi~.
\label{sq3}
\end{equation}
\twocolumngrid
To glean insight into Eq. (\ref{sq3}), note that if we take $\theta = 0$ or $\phi = 0$ as an example, 
the tangent of the complexity is proportional to the squeezing parameter. With say $\theta = 0$, $n=0$ and
for small values of $\phi$, we have ${\mathcal C} \sim 2\phi\xi_z$. Now, $\xi_z^2$ is also 
a measure of pairwise entanglement, with $\xi_z^2<1$ implying that the many-body density matrix is
not separable \cite{SDCZ}. Here we see that the operator complexity is proportional to the entanglement
measure of the system, for some specific choices of the rotation angles. 
Before closing this section, we point out that we can also write down the expression for the complexity in terms of the 
pairwise correlation function \cite{Review}
\begin{equation}
G_{1\mathbf{n},2\mathbf{n}}=\frac{1}{N-1}\Big[\frac{4}{N^{2}}\Big(N\big<
\big(\Delta \textbf{J}_{\mathbf{n}}(t)\big)^{2}\big>+\big<J_{\mathbf{n}}\big>^{2}\Big)-1\Big]~.
\end{equation}
Assuming $\theta=0$ we then have 
\begin{equation}
    \mathcal{C}\big(\phi,t\big)=2\arctan\bigg[\phi\sqrt{(N-1)G_{1z,2z}+1}\bigg]~~.
\end{equation}

\section{Complexity in the Tait-Bryan parametrisation}
\label{sec5}

Computation of the Nielsen complexity in the Euler angle parametrisation might not always be useful. For example, 
in the Lipkin-Meshkov-Glick model (to be elaborated upon shortly), it can be checked that during 
the time evolution, $\mathbf{J}_{y}(t)$ not only has a component along  $\mathbf{J}_{y}$ but also along
$\mathbf{J}_{x}$. This causes a problem in writing the rotation operator at an arbitrary time $t$,
because the general unitary rotation matrix of Eq. (\ref{Eadis-t}) does not have any 
$\mathbf{J}_{x}$ component in the lowest order of rotation angles.
Whereas this problem can still be dealt with by using the Euler angles parametrisation, but the price we pay is that 
the angles themselves become function of time and satisfy coupled differential equations \cite{ZGS,ZYS}. 
The solutions for these equations are difficult to obtain analytically even with simple functional forms of the 
time dependent frequency.  

Here we establish that this problem can be circumvented by using a different parametrisation of the 
general rotation operator, known in the literature as the Tait-Bryan (TB) angles \cite{HG}. 
In that case, instead of rotations only along fixed $y$ and $z$ axis, a general rotation is written in terms of 
a single rotation about all of the three fixed axis, i.e., the rotation operator now is given by 
\begin{equation}
U(\alpha,\beta,\gamma)=\exp\Big(-i\alpha \mathbf{J}_{x}\Big)\exp\Big(-i\beta \mathbf{J}_{y}\Big)\exp\Big(-i\gamma \mathbf{J}_{z}\Big)~.
\end{equation}
Note that the last rotation is around the fixed $x$ axis instead of $z$. Also, 
the TB rotations angles are denoted in this section by $\alpha, \beta, \gamma$ for convenience, and should 
not be confused with the Euler  angles used earlier.

As before the matrix form of the operator $U(\alpha,\beta,\gamma)$ in a preferred basis of 
the generators is given by 
\begin{equation}\label{TBUn}
U(\alpha,\beta,\gamma)=\left(\begin{array}{ccc}
1 &-\gamma& \beta \\
\gamma & 1 &-\alpha \\ 
-\beta & \alpha & 1\\
\end{array}
\right)~.
\end{equation}
The general unitary operator now has three independent angles and thus the unitary geometry is three dimensional. 
The line element of this geometry obtained by methods similar to the ones that we have used before is
lengthy and we omit it for brevity, while noting that on the $\alpha=0$ hypersurface, it reduces to 
the one in Eq. (\ref{bgmetric}). As before, we find that there is a hidden symmetry here, and that 
the metric can be put into a diagonal form by the writing it in terms of the new coordinates 
$(\rho, \Theta, \Phi)$ given in in terms of TB angles by the relations
\begin{equation}
\begin{split}
&\alpha\big(\rho,\Theta,\Phi\big)=\rho\sin\Phi~,~\beta\big(\rho,\Theta,\Phi\big)=\rho\sin\Theta\cos\Phi~,\\
&\gamma\big(\rho,\Theta\big)=\rho\cos\Theta\cos\Phi~~.
\end{split}
\end{equation}
In this new coordinate system, the metric acquires a simple diagonal form
\begin{equation}\label{TBdiagonal}
ds^{2}=\frac{1}{\big(1+\rho^{2}\big)^{2}}\bigg[4\text{d}\rho^{2}+\rho^{2}\big(\rho^{4}+5\rho^{2}+4\big)\Big(\text{d}\Phi^{2}+\cos^{2}\Phi \text{d}\Theta^{2}\Big)\bigg]~.
\end{equation}
Once again the coordinate $\Theta$ is cyclic, making the computations simple. 

We want to calculate the circuit complexity of creating the operator of $\mathcal{D}^{j}\big(\theta,\phi,t\big)$  
from the identity operator. However, the  matrix form of this operator, given by  
\begin{equation}\label{TBdis-t}
\mathcal{D}^{j}\big(\theta,\phi,t\big)=\left(
\begin{array}{ccc}
1 &-f_{3}(\theta,\phi,t) & f_{2}(\theta,\phi,t) \\
f_{3}(\theta,\phi,t) & 1 &-f_{1}(\theta,\phi,t) \\ 
-f_{2}(\theta,\phi,t) & f_{1}(\theta,\phi,t) & 1\\
\end{array}
\right)~
\end{equation}
now depends on three unknown functions $f_1, f_2, f_3$, whose form is to be determined by the Hamiltonian. 
To calculate this complexity we have to solve the geodesic equation for the metric in Eq. (\ref{TBdiagonal}) which satisfy the 
following boundary conditions at the two end points :
\begin{eqnarray}
\label{bctba}
&~&\alpha\big(\tau=0\big)=\beta\big(\tau=0\big)=\gamma\big(\tau=0\big)=0~,\nonumber\\
&~&\alpha \big(\tau=1\big)=f_{1}(\theta,\phi,t)~,~
\beta\big(\tau=1\big)=f_{2}(\theta,\phi,t)~,\nonumber\\ 
&~&\gamma\big(\tau=1\big)=f_{3}(\theta,\phi,t)~.
\end{eqnarray}
Since $\Theta$ is a cyclic coordinate we have for the conserved quantity $L$ the following first order equation for $\Theta$
\begin{equation}
\label{thetafirst}
\dot{\Theta}=\frac{L \big(1+\rho^{2}\big)}{2\rho^{2}\big(\rho^{2}+4\big)\cos^{2}
{\Phi}}~.
\end{equation}
and  the equations for $\Phi(\tau)$ and $\rho(\tau)$ are now given by
\begin{equation}
\label{phirhodot}
\begin{split}
&\frac{\text{d}}{\text{d}\tau}\left[\frac{2\rho^{2}\big(\rho^{2}+4\big)}{\big(1+\rho^{2}\big)}
\dot{\Phi}\right]-\frac{2\rho^{2}\big(\rho^{2}+4\big)}
{\big(1+\rho^{2}\big)}\cos{\Phi}\sin{\Phi}\dot{\Theta}^{2}=0~,\\
&\dot{\rho}^{2}=\frac{K^{2}\big(1+\rho^{2}\big)^{2}}{4}-\frac{L^{2}\big(1+\rho^{2}\big)^{3}}
{16\rho^{2}\big(\rho^{2}+4\big)\cos^{2}
{\Phi}}\\
&-\frac{\rho^{2}\big(\rho^{2}+4\big)\big(1+\rho^{2}\big)}{4}\dot{\Phi}^{2}~,
\end{split}
\end{equation}
where $K^{2}=g_{ij}\dot{y}^{i}\dot{y}^{j}$. 
It is difficult to eliminate $\dot{\Phi}$ from Eq. (\ref{phirhodot}) to obtain a first order equation 
for $\rho(\tau)$ only. Instead of directly solving these equations, we will use the well known Hamilton-Jacobi method to 
separate these. For a geodesic $y^{i}(\tau)$ of the metric $g_{ij}$ the Hamilton-Jacobi (HJ) equation for the
Hamilton principal function $S$ is defined by (the method is standard and is routinely used in 
general relativity, and can be found in many textbooks, see, e.g., the authoritative account in \cite{SC})
\begin{equation}
2\frac{\partial S}{\partial \tau}=g^{ab}\frac{\partial S}{\partial y^{a}}\frac{\partial S}{\partial y^{b}}~.
\end{equation}
For our case the HJ equation is given by
\begin{equation}
\begin{split}
2\frac{\partial S}{\partial \tau}&=\frac{\big(1+\rho^{2}\big)^{2}}{4}\left(\frac{\partial S}
{\partial \rho}\right)^{2}
+\frac{\big(1+\rho^{2}\big)}{\rho^{2}\big(\rho^{2}+4\big)}\left(\frac{\partial S}
{\partial \Phi}\right)^{2}\\
&+\frac{\big(1+\rho^{2}\big)}{\rho^{2}\big(\rho^{2}+4\big)\cos^{2}\Phi}
\left(\frac{\partial S}{\partial \Theta}\right)^{2}~.
\end{split}
\end{equation}
Since we already know that the momentum associated with $\Theta$ and $g_{ab}\dot{y}^{a}\dot{y}^{b}$ are 
two constants of motion, we assume that a solution $S$ of the HJ equation  can be 
written in the following separable form
\begin{equation}
S=\frac{1}{2}K^{2}\tau+L\Theta +S_{\Phi}(\Phi)+S_{\rho}(\rho)~.
\end{equation}
As indicated, $S_{\rho}(\rho)$ and $S_{\Phi}(\Phi)$ are two functions of their single arguments. 
Substituting $S$ in the HJ equation we have
\begin{equation}
\begin{split}
K^{2}&=\frac{\big(1+\rho^{2}\big)^{2}}{4}\left(\frac{\text{d}S_{\rho}}{\text{d}\rho}\right)^{2}+\frac{\big(1+\rho^{2}\big)}
{\rho^{2}\big(\rho^{2}+4\big)}\left(\frac{\text{d}S_{\Phi}}{\text{d}\Phi}\right)^{2}\\
&+\frac{L^{2}\big(1+\rho^{2}\big)}
{\rho^{2}\big(\rho^{2}+4\big)\cos^{2}\Phi}~.
\end{split}
\end{equation}
After a bit of algebraic manipulation, we can rewrite this as
\begin{equation}
\begin{split}
&\frac{K^{2}\rho^{2}\big(\rho^{2}+4\big)}{\big(1+\rho^{2}\big)}-\frac{1}{4}\rho^{2}\big(\rho^{2}+4\big)\big(\rho^{2}
+1\big)\left(\frac{\text{d}S_{\rho}}{\text{d}\rho}\right)^{2}\\
&=L^{2}\sec^{2}\Phi +\left(\frac{\text{d}S_{\Phi}}{\text{d}\Phi}\right)^{2}~.
\end{split}
\end{equation}
Now the left side of this equation is solely a function of $\rho$, and the right side is a function only of $\Phi$. Thus 
to satisfy this equation, both side must be equal to a constant which we call $\mathcal{M}$.

Now that we have the separated the HJ equation, we can get the first order equations in the coordinates by 
using the formula
\begin{equation}
p_{a}=g_{ab}\frac{\text{d}y^{a}}{\text{d}\tau}=\frac{\partial S}{\partial y^{a}}~,
\end{equation}
and we find that these are given by 
\begin{equation}
\begin{split}
&\frac{\text{d}\rho}{\text{d}\tau}=\frac{1}{2}\sqrt{\frac{\big(1+\rho^{2}\big)^{3}}{\rho^{2}\big(\rho^{2}+4\big)}}
\sqrt{\frac{K^{2}\rho^{2}\big(\rho^{2}+4\big)}{\big(1+\rho^{2}\big)}-\mathcal{M}}~,\\
&\frac{\text{d}\Phi}{\text{d}\tau}=\frac{\big(1+\rho^{2}\big)}{\rho^{2}\big(\rho^{2}+4\big)}\sqrt{\mathcal{M}-L^{2}\sec^{2}\Phi}~.
\end{split}
\end{equation}
We need to solve these two equations along with the first order equation for $\Theta$ (Eq. (\ref{thetafirst})),  with the boundary conditions mentioned in Eq. (\ref{bctba}). 
For this we first notice in order to avoid divergences at $\rho=0$ and hence to be compatible with the boundary conditions, we need to put $L=\mathcal{M}=0$. Then, 
we obtain, 
\begin{equation}
\Theta=\Theta_{0}~,~~\Phi=\Phi_{0}~,~~\rho\big(\tau\big)=\tan\Big[\frac{K\tau}{2}-C_{3}\Big]~,
\end{equation}
where $\Theta_{0}$ and $\Phi_{0}$ are constants. 
Note that the solution for $\rho$ is the same as obtained before for the case of ordinary Euler angle parametrisation. 
However, the expression for $K$ is different from the previous case due to the presence of the extra coordinate.
To see this, we first note that the boundary conditions here imply that 
\begin{equation}
\Theta_{0}=\arctan\Big[f_{2}/f_{3}\Big]~,~~
\Phi_{0}=\arctan\bigg[\frac{f_{1}}{\sqrt{f_{2}^{2}+f_{3}^{2}}}\bigg]~.
\end{equation}
These then give our final expression for the complexity, 
\begin{equation}\label{T-Bcomplexity}
K=2\arctan\Bigg[\sqrt{f_{1}^{2}+f_{2}^{2}+f_{3}^{2}}\Bigg]+2n\pi~.
\end{equation}

\subsection{Class-3 : The Lipkin- Meshkov-Glick model with TB parameters}
\label{class3}

As a concrete example, we consider the  Lipkin- Meshkov-Glick (LMG) model which describes the 
interaction of $N$ spin $1/2$ particles in presence of an external magnetic field \cite{LMG}. 
When the magnetic field is applied along the $z$ direction, the Hamiltonian of the LMG model can be written as \cite{JS,LYC}
\begin{equation}
\mathbf{H}_{LMG}=-\frac{\lambda}{N}\left(\mathbf{J}_{x}^{2}+\kappa\mathbf{J}_{y}^{2}\right)-B \mathbf{J}_{z}~.
\end{equation}
Here, as before, $\mathbf{J}_{i}$ is the total spin operator in the $i$th direction. 
The parameter $\lambda>0$ characterises the strength of the interaction between the spins, and $0\leq\kappa\leq1$ 
characterises the anisotropy of this interaction. 

The Heisenberg equations of motion are given by
\begin{equation}\label{HeLMG}
\begin{split}
&\frac{d\mathbf{J}_{x}(t)}{dt}=-\frac{2\kappa}{N} \mathbf{J}_{(y}(t)\mathbf{J}_{z)}(t)+B\mathbf{J}_{y}(t)~,~~ \\
&\frac{d\mathbf{J}_{y}(t)}{dt}= \frac{2}{N}\mathbf{J}_{(x}(t)\mathbf{J}_{z)}(t)-B \mathbf{J}_{x}(t)~,~~\\
&\frac{d\mathbf{J}_{z}(t)}{dt}=-\frac{2\lambda \big(1-\kappa\big)}{N}\mathbf{J}_{(x}(t)\mathbf{J}_{y)}(t)~.
\end{split}
\end{equation}
Since the general solution of these equations are difficult to obtain, we shall focus here on
some simple cases. First consider the case of isotropic LMG model, characterised by $\kappa=1$ \cite{JS}. 
In that case, the Hamiltonian commutes with $\mathbf{J}_{z}$ and hence is diagonal in the $\big|J,m\big>$ 
representation. The energy eigenvalues are then given by 
\begin{equation}
E=-\frac{\lambda}{N}\left(J(J+1)-m^{2}\right)-Bm~,
\end{equation}
here $J$ now denotes the eigen value of the collective spin operator. This also means that $\mathbf{J}_{z}(t)$ is time independent, which can be seen from the last relation of 
Eq. (\ref{HeLMG}). Since $\mathbf{J}_{z}(t)$ is time independent, the other two equations of motion are easy to solve, 
and are given by the periodic functions 
\begin{equation}
\begin{split}
\mathbf{J}_{x}(t)= \mathbf{J}_{x}\cos \big[\omega_{1}t\big]+\mathbf{J}_{y}\sin \big[\omega_{1}t\big]~\\
\mathbf{J}_{y}(t)= \mathbf{J}_{y}\cos \big[\omega_{1}t\big]-\mathbf{J}_{x}\sin \big[\omega_{1}t\big]~,
\end{split}
\end{equation}
with $\omega_{1}=B+1$, where for convenience we have set $\mathbf{J}_{z}=-N/2$ and as before when the argument of the angular momentum operator is not explicitly written it indicates value at $t=0$.  Now the time evolution of the operator $\mathcal{D}^{j}\big(\theta,\phi\big)$ is obtained to be
\begin{equation}\label{Dtptlmg}
\begin{split}
&\mathcal{D}^{j}\big(\theta,\phi,t\big)=
\exp\Big(-i\phi \mathbf{J}_{z}\Big)\exp\Big(-i\theta \Big\{\mathbf{J}_{y}\cos \big[\omega_{1}t\big] \\
&-\mathbf{J}_{x}
\sin \big[\omega_{1}t\big]\Big\}\Big)=\exp\Big(-i\big[\phi+\omega_{1} t\big] \mathbf{J}_{z}\Big)\exp\Big(-i\theta \mathbf{J}_{y}\Big)~.
\end{split}
\end{equation}
Here, in the last step, we have employed the transformation from Euler angle to the axis angle representation described in Appendix \ref{AppendixA}.
As can be seen, the operator complexity in this case is the same as the spin magnet interaction.

Now we shall consider a more general case, namely the anisotropic LMG model in the frozen spin approximation. 
As usual in the frozen spin approximation, we assume that under the influence of the external magnetic field, the spin 
component along that direction i.e., $\mathbf{J}_{z}$ remains fixed. Once again, we take $\mathbf{J}_{z}=-N/2$. 
This approximation is valid when the strength of the external field is much greater than the spin-spin interaction. 
Then the solutions of the equation of motion are  given by (see, e.g., \cite{LYC})
\begin{equation}
\begin{split}
&\mathbf{J}_{y}(t)= \mathbf{J}_{y}\cos \big[\omega_{B}t\big]-\sqrt{\frac{B+1}{B+\kappa}}\mathbf{J}_{x}\sin \big[\omega_{B}t\big]~,\\
&\mathbf{J}_{x}(t)= \mathbf{J}_{x}\cos \big[\omega_{B}t\big]+\sqrt{\frac{B+\kappa}{B+1}}\mathbf{J}_{y}\sin \big[\omega_{B}t\big]~,
\end{split}
\end{equation}
with $\omega_{B}=\sqrt{\big(B+1\big)\big(B+\kappa \big)}$.
Notice that these solutions reduce to that of 
the isotropic case when $\kappa=1$, because in both the cases, $\mathbf{J}_{z}(t)$ remains constant. In the isotropic 
case the solutions become exact.

As before,  the operator $\mathcal{D}^{j}\big(\theta,\phi,t\big)$  is given by 
\begin{equation}\label{DtptLMG}
\begin{split}
&\mathcal{D}^{j}\big(\theta,\phi,t\big)=
\exp\Big(-i\phi \mathbf{J}_{z}\Big)\exp\Big(-i\theta \Big\{\mathbf{J}_{y}\cos 
\big[\omega_{B}t\big]\\
&-\sqrt{\frac{B+1}{B+\kappa}}\mathbf{J}_{x}\sin \big[\omega_{B}t\big]\Big\}\Big).
\end{split}
\end{equation}
Now a comparison with previous cases (for example Eq. (\ref{Dtssf})) show that since 
here  $\mathbf{J}_{y}(t)$ has a nonzero component along $\mathbf{J}_{x}$ it gives a 
contribution in the first order of $\theta$.
If we work with the  unitary operator in the Euler angle representation of previous sec-\ref{sec3}, 
then it  does not have any rotation about $x$ axis. The same situation arose previously in the case of 
spin-magnet interaction as well as in the isotropic LMG model discussed previously. 
However in both the cases, the problem was solved by using the equivalence between  the axis-angle and 
Euler angle representation  described in Appendix \ref{AppendixA}. Since the coefficient of $\mathbf{J}_{x}\sin \big[\omega_{B}t\big]$ is not unity, a 
little inspection shows that the same trick can not be used here.
The  advantage of the TB parametrisation used in this section over the standard  Euler angles is apparent 
in this case - a rotation along $x, y$ or $z$ axis can be incorporated with TB angles not with the Euler ones 
(at least when the angles themselves are time independent).

To calculate the complexity of LMG model at time $t$ we simply write down the matrix form of 
Eq. (\ref{Dtptlmg}), and compare it with that 
of Eq. (\ref{TBdis-t}) of the TB parametrisation. We find, 
\begin{equation}\label{LMGfs}
f_{1} =-\theta \sqrt{\frac{B+1}{B+\kappa}}\sin \big[\omega_{B}t\big]~,~f_{2}=\theta \cos\big[\omega_{B}t\big]~,~f_{3}=\phi~.
\end{equation}
These  can be readily used in the general expression Eq. (\ref{T-Bcomplexity}) to obtain the complexity. 

\section{Conclusions}
\label{sec6}

In this paper, we have studied Nielsen's complexity in the context of coherent spin state operators. We have
constructed a simple formula for the complexity for linearised operators that correspond to preparing 
a target state from a given reference state. This was achieved by constructing the metric on the space of unitaries of
the rotation operator by employing the Euler angle representation, 
and computing analytic solutions of the resulting geodesics with appropriate boundary
conditions. We considered three  models, with increasing levels of complication. Our first example was the
ubiquitous spin-magnetic field coupling. This was followed by the one-axis twisting model where we showed an
interesting possibility of relating the complexity with pairwise entanglement. Finally we considered the
LMG model, where the usual Euler angle parametrisation was found to be less useful and we resorted to a
Tait-Bryan parametrisation. 

It is useful to contrast the results presented here with the Fubini-Study complexity (see, e.g. \cite{tapo1}). This is
an alternative characterisation of complexity where one works directly with the metric on the parameter
manifold of quantum states. In the time-independent case, the parameter manifold of the CSS has been worked
out in \cite{ProvostVallee} and corresponds to the two-sphere. The geodesics then are simply great circles
on this sphere and are different from the ones on the space of unitaries that we have computed for the 
CSS operators. 

In this paper we gave indication that the Nielsen complexity of the CSS operators might be related to pairwise
entanglement. It would be very interesting to quantify this further. 

\appendix

\section{}
\label{AppendixA}

Here we detail some details of the results presented in subsection \ref{class1}. 
At time $t$ the expression for SCS of the Hamiltonian  $\mathbf{H}_{sm}=-\mathbf{S}_{z}  \mathbf{B}$ is given by 
(here $j$ in eq.(\ref{CSS}) are $s$, the eigen values of total spin operator $\mathbf{S}$)
\begin{equation}
\begin{split}
&\big|\theta,\phi,t\big>
=\sum_{m=-s}^{s}\sqrt{\left(
\begin{array}{ccc}
2s  \\
s+m\\
\end{array}
\right)}\Big(\cos \frac{\theta}{2}\Big)^{s+m}\Big(\sin \frac{\theta}{2}\Big)^{s-m}\times\\
&e^{i\big(s-m\big)\phi}e^{imBt}\big|J,m\big>
= e^{-iBts}\big|\theta,\big(\phi+Bt\big)\big>.
\end{split}
\end{equation}
We can calculate $\mathcal{D}^{j}\big(\theta,\phi,t\big)$ directly in the Heisenberg picture to be
\begin{equation}
\label{dtpt}
\begin{split}
&\mathcal{D}^{j}\big(\theta,\phi,t\big)=\exp\Big(-i\phi \mathbf{S}_{z}\Big)
\exp\Big(-i\theta \mathbf{S}_{y}(t)\Big)\\
&=\exp\Big(-i\phi \mathbf{S}_{z}\Big)\exp
\Big(-i\theta \Big\{\mathbf{S}_{y}\cos\big[Bt\big]-\mathbf{S}_{x}\sin\big[Bt\big]\Big\}\Big)~.
\end{split}
\end{equation}
Note that this form of the operator $\mathcal{D}^{j}\big(\theta,\phi,t\big)$ is the same as the one at $t=0$ with 
$\phi$ replaced by $\big(\phi+Bt\big)$. To see this, we compare the second exponential operator in the expression 
above with the rotation operator used to construct a CSS in the axis-angle representation of eq.(\ref{diaxa}).
Thus the second exponential operator in eq.(\ref{dtpt}) is nothing but the operator which rotates the unit vector 
$\mathbf{z}$ along the $z$ direction  by an angle $\theta$ with respect to the unit vector 
$\mathbf{n}_{\phi}=\{-\sin\big[Bt\big] , \cos\big[Bt\big],0 \}$, albeit written in the axis-angle representation. 
But we know how to represent the same rotation in terms of Euler angles, which is given by the rotation operator 
$\mathcal{R}(\phi=Bt,\theta,0)=\mathcal{R}_{z}(\phi=Bt)\mathcal{R}_{y}(\theta)$. Thus we finally have the operator at an 
arbitrary time to be 
\begin{equation}
\begin{split}
&\mathcal{D}^{j}\big(\theta,\phi,t\big)=\exp\Big(-i\phi \mathbf{S}_{z}\Big)\exp
\Big(-iBt \mathbf{S}_{z}\Big)\exp\Big(-i\theta \mathbf{S}_{y}\Big)~\\
&=\exp\Big(-i\big[\phi+Bt\big] 
\mathbf{S}_{z}\Big)\exp\Big(-i\theta \mathbf{S}_{y}\Big)~.
\end{split}
\end{equation}
Now taking linear form of this operator for small values of the rotation angle, the complexity at an time $t$ is 
straightforwardly obtained as Eq. (\ref{class1eq}).


\end{document}